\documentclass[a4paper,12pt]{article}

\begin{document}
%================

\begin{center}
{\Large
Implementation of the Non-Linear Gauge into GRACE
\footnote
{Talk presented by K.Kato at
6-th AIHENP, April 1999,
University of Crete.}
\\
}

\vspace{8mm}

{\large G. B\'elanger${}^{1)}$, F. Boudjema${}^{1)}$,
J.Fujimoto${}^{2)}$, T.Ishikawa${}^{2)}$, \\ T. Kaneko${}^{3)}$,
K. Kato${}^{4)}$, V. Lafage${}^{2)}$, N. Nakazawa${}^{4)}$, Y.
Shimizu${}^{2)}$ \\ %and Minami-Tateya Collaboration \\

\vspace{4mm}

{\it 1) LAPTH,  Annecy-le-Vieux F-74941, France} \\ {\it
2) KEK, Tsukuba, Ibaraki 305--0801, Japan} \\ {\it 3)
Meiji-Gakuin University, Totsuka, Yokohama 244--0816, Japan} \\
{\it 4) Kogakuin University,  Shinjuku, Tokyo
163--8677, Japan} \\

\vspace{10mm}
}

\end{center}

\noindent
{\bf Abstract}
\begin{quote}
A general non-linear gauge condition is implemented into {\tt
GRACE}, an automated system for the calculation of physical
processes in high-energy physics. This new gauge-fixing is used as
a  very efficient means to check the results of  large scale
evaluation in the standard model computed automatically. We
report on some systematic test-runs which have been performed  for
one-loop two-to-two processes to show the validity of the gauge
check.
\end{quote}

\vspace{8mm}

A major part of the theoretical predictions in  high-energy
physics is based on perturbation theory. However the complexity
increases rapidly as one moves to higher orders in perturbations,
like when dealing with loop corrections or when dealing with
many-body final states. In many instances calculations,
if done by hand, become intractable and prone to error.  Since
perturbation theory is a well-established algorithm, it is
possible to construct a system or software to perform these
calculations automatically.

There are several systems operating as expert systems for
high-energy physics\cite{review}. The {\it Minami-Tateya}\, group
has developed the system named {\tt GRACE}.\cite{grace} Its
structure is depicted in the figure. The system can,
in principle, deal with the perturbative series up to any order. For
instance all diagrams contributing to a process are generated
automatically given the order of perturbation, and specifying the
particles. However, due to the handling of the  loop integrals,
practical calculations are, for the moment, restricted to tree and
one-loop orders. Two-loop calculation is possible only for some
limited cases. The {\tt GRACE} system can work for any type of
theory, once the model file is
implemented. Here, the model file is a database which  stores all
component fields and interactions contained in the Lagrangian of
the theory.  Besides the model file, peripheral parts are
sometimes required.  For instance, the structure of vertex
in the theory is absent in the tool box of {\tt GRACE},
the definition should be added. 
Also, soft correction factor, kinematics code, the interface
to structure functions, parameter control section, and so forth 
might be supplied if necessary. The system is
versatile enough to include new features and such added 
components would become a part of the new version of {\tt GRACE}.
This report is confined to the
calculation based on the so-called standard model. The extension
to SUSY is presented in a separated talk.\cite{fujimoto}

In contrast to the manual computation, the theoretical prediction
from an automated system is obtained by invoking several commands
at a terminal with some elapsed time ( which is often rather long
), due for instance to numerical  integrations over phase-space.
Especially with an automated system it is difficult to judge the
reliability of the final result, hence a need for  a built-in
automated function to confirm the results. At tree-level, the
check of gauge invariance has been  shown to be powerful. Within
{\tt GRACE}, the comparison is done between the unitary gauge and
the covariant gauge and we observe $\sim 15$($\sim 30$)
digits agreement in double(quadruple) precision computation.

%==============================
\begin{figure}[htbp]
\input{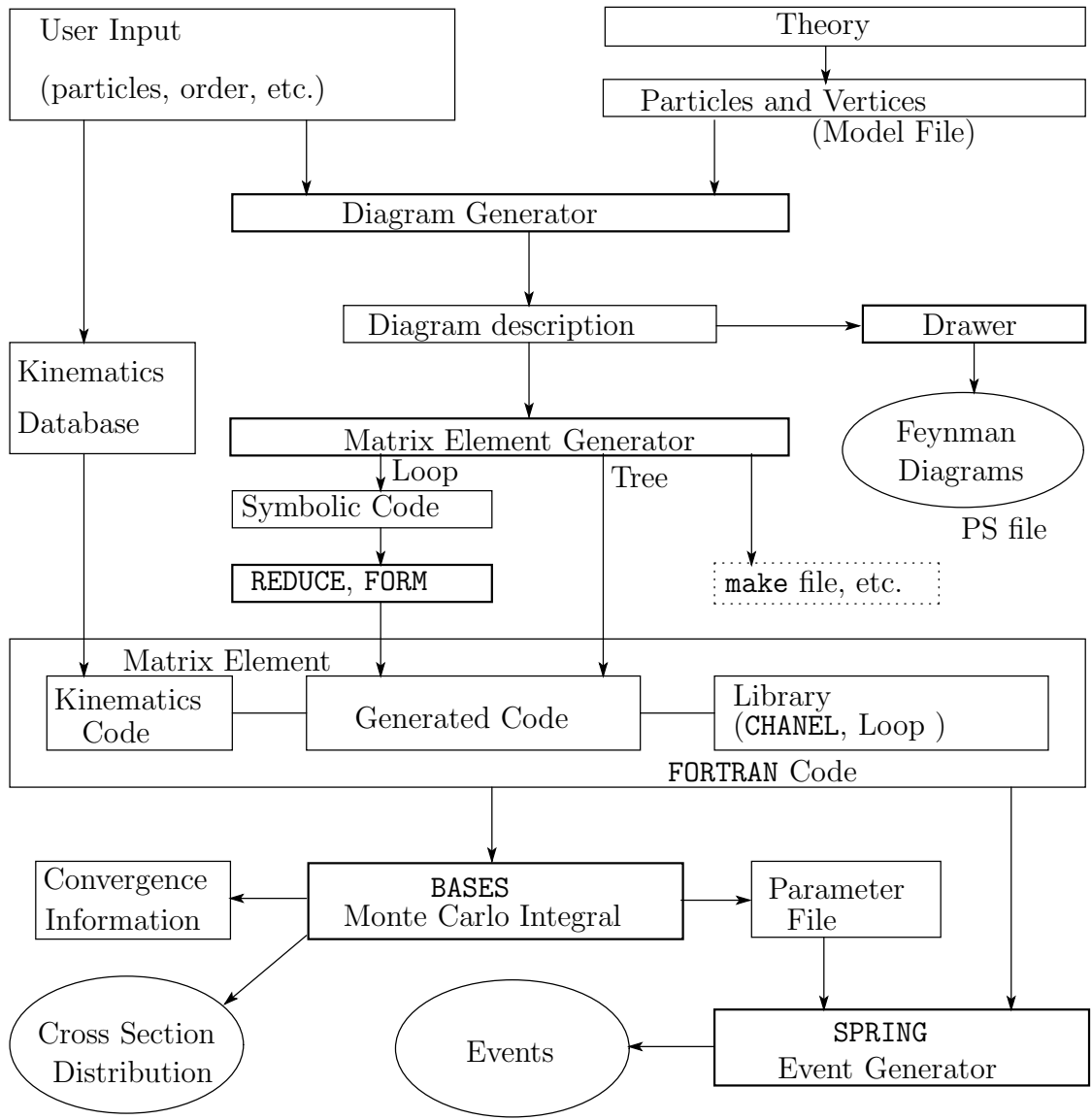}
        \caption{{\tt GRACE} System Flow}
\end{figure}
%==============================
\clearpage

In the one-loop case, the calculation in the
unitary gauge does not work well. 
Within {\tt GRACE} one of the problems in this gauge is that the library
containing the various loop integrals  is designed assuming that
the numerator of vector particles is $g^{\mu\nu}$. For instance,
the library for vertex functions is implemented for the numerator
of 3rd order polynomial in the loop momentum, while the handling
of 9th order one is required in the unitary gauge. This not only
creates very large expressions but also introduces terms with
large superficial divergences that eventually need to be canceled
precisely between many separate contributions. 

The non-linear
gauge\cite{nlg0} is introduced to make the gauge-check possible
within {\tt GRACE}. We take a generalized non-linear gauge fixing
condition for the standard model\cite{nlg}:
\begin{equation}
\begin{array}{rl}
{\cal L}_{\rm GF}=& \displaystyle{
-\frac{1}{\xi_W}\left|(\partial_{\mu}-ie\tilde{\alpha}A_{\mu}
-ig\cos\theta_W\tilde{\beta}Z_{\mu})W^{+\mu}
+\xi_W\frac{g}{2}(v+\tilde{\delta}H+i\tilde{\kappa}\chi_3)\chi^{+}\right|^2
}\\
{ } & { } \\
{ } & \displaystyle{
-\frac{1}{2\xi_Z}\left(\partial_{\mu}Z^{\mu}
+\xi_Z\frac{g}{2\cos\theta_W}(v+\tilde{\epsilon}H)\chi_3\right)^2
\quad -\frac{1}{2\xi_A}(\partial_{\mu}A^{\mu})^2}
\end{array}
\end{equation}%
We take $\xi_W=\xi_Z=\xi_A=1$ so that the numerator structure of
vectors is $g^{\mu\nu}$ as with the usual 'tHooft-Feynman gauge.
Any of the other parameters can then provide a gauge check. Then
the {\tt GRACE} library to compute loop amplitudes works without
any change. When
$\tilde{\alpha}=\tilde{\beta}=\tilde{\delta}=\tilde{\kappa}=
\tilde{\epsilon}=0$, the gauge turns to the 'tHooft-Feynman gauge.
For instance, when $\tilde{\alpha}=1$
($\tilde{\beta}=-\tan^2\theta_w$), $W^{\pm}\chi^{\mp}\gamma$
($W^{\pm}\chi^{\mp}Z$) vertex disappears. Though in the non-linear
gauge we have new vertices in the ghost sector, e.g.,
ghost-ghost-vector-vector vertices, we can reduce the total number
of diagrams and therefore speed up the calculation of many
processes. Of course the Feynman rules in the model file is
revised to take into account the modifications due to the
non-linear gauge. This includes all the vertices at tree-level as
well as all appropriate counter terms needed for any 1-loop
calculation. As a check on the new input file, we have
confirmed that the results is unchanged from the original standard
model case when all non-linear gauge parameters are 0.

We have tested the system for several tree-level processes as well
as  one-loop two-to-two processes. Here we report on the tests
done at the 1-loop level. The ultraviolet divergence is
regularized by dimensional regularization where space-time
dimension is $n=4-2\varepsilon$. In the code generated by {\tt
GRACE}, the divergence is kept as a variable {\tt Cuv} which
stands for $1/\varepsilon - \gamma_E + \log 4\pi$. As a first step
to check the system, we have compared, for a given random set of
the gauge parameters, the results with letting {\tt Cuv}$=0$ and
that with {\tt Cuv}$\ne 0$. Since the agreement is exact, the
system passes the first diagnosis, renormalizability. Then, we
proceed to check that the finite result  is independent of the
choice of set for the non-linear gauge parameters.

Some of the processes are listed in the table  below. The
center-of-mass energy and the masses used in the computation are
as follows: $ W=500 {\rm GeV} $, $ M_Z=91.187{\rm GeV},
  M_W=80.37{\rm GeV},
  M_H=100{\rm GeV},
  M_t=174{\rm GeV}$. 
For the regularization of the 
infra-red divergences,
we introduce a fictitious photon mass, and its value
in the computation is 
  $\lambda=10^{-6}{\rm GeV}. $
In the table, we present the value of
\[
\left(
  \frac{d \sigma^{1-loop}}{d \cos\theta}
  -\frac{d \sigma^{tree}}{d \cos\theta}
\right)_ {\theta=10^{\circ}(\cos\theta=0.985)} \quad\propto\quad 2
\Re \left(T^{loop}\cdot T^{tree\ \dagger}\right), $$ 
and LG
stands for the linear gauge('tHooft-Feynman gauge) case with
{\tt Cuv}=0 and NLG does for the non-linear gauge case with
{\tt Cuv}=1, $\tilde{\alpha}=\tilde{\beta}=1$,
$\tilde{\delta}=\tilde{\kappa}=\tilde{\epsilon}=0$. The latter
case corresponds to the background gauge
formalism.\cite{background} The number of diagrams depends on the
choice of the gauge-fixing condition since some vertices are not
present in some gauges. The counting of the total number of
diagrams in the Table refers to all possible diagrams and
therefore with appropriate choices of gauge this number may be
reduced. The number of diagrams involving counter-terms insertions
and that with self-energy contributions is denoted by CT and SE,
respectively.

\vspace{4mm}

\begin{tabular}{lrrrcc}
\hline
           & \multicolumn{3}{c}{Number of Diagrams}
 &\multicolumn{2}{c}{$d\sigma/d\cos\theta$[pb]} \\
\hline
process &   tree&    1-loop&(CT:SE)&    LG &    NLG \\
\hline
\rule{0mm}{5mm}
$e^+e^- \rightarrow t\bar{t} $
 &   2 &      52&  (4 : 6)&    -0.870876519  & -0.87087619 \\
\rule{0mm}{5mm}
$e^+e^- \rightarrow HZ $
 &   1 &     119&  (3 : 4)&    -0.03174046785& -0.03174046785 \\
\rule{0mm}{5mm}
$e^+e^- \rightarrow W^-W^+ $
 &   3 &     152&  (5 : 5)&    -0.9963368092 & -0.9964451605 \\
\hline
\rule{0mm}{5mm}
$t\bar{b} \rightarrow t\bar{b} $
 &   6 &     268&  (12:14)&     33.75029132  &   33.75132514 \\
\rule{0mm}{5mm}
$W^+W^- \rightarrow t\bar{t} $
 &   4 &     238&  (8 : 6)&    -0.08938607492&  -0.08938607286 \\
\rule{0mm}{5mm}
$ZH \rightarrow t\bar{t} $
 &   4 &     352&  (8 : 8)&     2.672194263  &      2.672194265 \\
\hline
\rule{0mm}{5mm}
$W^+\gamma \rightarrow t\bar{b} $
 &   4 &     238&  (8 : 6)&    -0.5998664910 &  -0.5998663896 \\
\rule{0mm}{5mm}
$W^+Z \rightarrow t\bar{b} $
 &   4 &     282&  (8 : 6)&    -0.2216982981 &   -0.2216940817 \\
\rule{0mm}{5mm}
$W^+H \rightarrow t\bar{b} $
 &   4 &     283&  (8 : 6)&     0.04785521591&    0.0478570236 \\
\hline
\end{tabular}

\vspace{4mm}

We note that the agreement  and hence the gauge independence of
the result is excellent. The accuracy is only limited by that of
the evaluation of the numerical loop integrals. Through detailed
inspections of individual diagrams we have noted that, expectedly,
the gauge independence requires different diagrams to combine in
order to produce a gauge-parameter independent result. In this
sense, the check by gauge invariance is a powerful diagnostics.
The comparison is still in progress for other processes at
one-loop to establish the validity of the system. We can then
proceed to more complicated processes, {\it i.e.}, one-loop
two-to-three or two-to-four processes and can confirm the results
of large scale computation by the non-linear gauge method
presented here.

%======================================================================
\vspace{10mm}

\noindent Acknowledgments

\begin{quote}
The authors would like to thank the local organizing committee
of AIHENP99 for the excellent organization.
They also thank the colleagues in {\it Minami-Tateya}\, group
and the CPP collaboration in Japan, France, and Russia.
This work is supported in part by the Ministry of Education,
Science and Culture, Japan under the
Grant-in-Aid for Scientific Research Program No.10640282
and No.09649384.

\end{quote}

%======================================================================

%======================================================================
\end{document}